\input xy
\xyoption{all}
\input epsf.tex

\documentclass[prd, onecolumn,superscriptaddress,altaffilletter,showpacs,nofootinbib]{revtex4}
\usepackage[dvips]{graphicx}
\usepackage{amsmath}

\newcommand{\eea}{\end{eqnarray}}

\begin{document}

\title{Spherically symmetric solution in a space-time with torsion}

\author{Filemon Farf\'an}
\email{keops_levy9@hotmail.com}
\affiliation{Centro Universitario de Ciencias Exactas e Ingenier\'ia, Universidad de
Guadalajara, Av. Revoluci\'on 1500, S.R., 44430,  Guadalajara, Jalisco, M\'exico.}
\author{Ricardo Garc\'ia-Salcedo}
\email{rigarcias@ipn.mx}
\affiliation{CICATA-Legaria, IPN, Av. Legaria 694, Col. Irrigaci\'on, CP.11500, M\'exico City, M\'exico. }
\author{Oscar Loaiza-Brito}
\email{oloaiza@fisica.ugto.mx}
\affiliation{Departamento de F\'isica, DCI,  Campus Le\'on, Universidad de Guanajuato, C.P. 37150, Guanajuato, M\'exico.}
\author{Claudia Moreno}
\email{claudia.moreno@cucei.udg.mx}
\affiliation{Centro Universitario de Ciencias Exactas e Ingenier\'ia, Universidad de
Guadalajara, Av. Revoluci\'on 1500, S.R., 44430,  Guadalajara, Jalisco, M\'exico.}
\author{Alexander Yakhno}
\email{alexander.yakhno@cucei.udg.mx}
\affiliation{Centro Universitario de Ciencias Exactas e Ingenier\'ia, Universidad de
Guadalajara, Av. Revoluci\'on 1500, S.R., 44430,  Guadalajara, Jalisco, M\'exico.}

\date{\today}

\begin{abstract}
By using the method of group analysis,  we obtain a new exact evolving and spherically symmetric solution of the Einstein-Cartan equations of motion,  corresponding to a space-time threaded with a three-form Kalb-Ramond field strength.  The solution describes in its more generic form, a space-time which scalar curvature vanishes for large distances and for large time.  In static conditions, it reduces to  a classical wormhole solution and to a  exact solution with a localized scalar field and a torsion kink, already reported in literature. In the process we have found evidence towards the construction of more new solutions.
\end{abstract}

\maketitle

\section{Introduction}

Unification of all fundamental forces in nature has represented a difficult task along the last few decades.  The most promising models are encoded within the context of string theory. For instance, there are models in  heterotic string theories in which grand unifications theories are present, while in type II superstring theories, partial unification is an available option in several model-dependent scenarios, where there exists the possibility to study geometric aspects of general relativity at the microscopic level (see \cite{Marchesano:2007de} and references therein).\\

In order to construct effective four-dimensional theories with a minimal (non-zero) number of supersymmetric generators, critical type II superstrings are typically compactified on Calabi-Yau (CY) threefolds, which break three-quarters of the total supersymmetry. Roughly speaking, this follows from the fact that, in the absence of other supergravity fields,  the ten-dimensional gravitino satisfies the constraint  $\delta\Psi_M^A=\nabla_M \epsilon^A=0$, implying the presence of two four-dimensional covariantly constant spinors.
Under these fluxless conditions, it is possible to construct a maximally symmetric four-dimensional theory (Minkowski, de Sitter or anti-de Sitter). However, fluxless compactification scenarios suffer from the moduli stabilization problem.\\

Moduli are stabilized, once one includes fluxes in the compactification process. Nevertheless, fluxes back-react and their presence modifies the internal geometry, forcing us to depart from the nice and smooth CY geometry. Even more, the presence of fluxes contributes to the ten-dimensional gravitino variation by
\begin{eqnarray}
\delta\Psi_M^A=\nabla_M\epsilon^A+ \kappa_M\epsilon^A,
\end{eqnarray}
where $\kappa_M$ encodes the flux contributions: it vanishes in the fluxless case. Notice however, that to preserve two supersymmetries in the effective four-dimensional theory, the variation of the gravitino must vanish, implying that the four-dimensional component of the supersymmetric parameter is not covariantly constant with respect to the Levi-Civita equation, but with respect to a connection with torsion. It is then clear that,  the effective four-dimensional supergravity theories obtained from the compactification process,  must be different from the theories constructed in the fluxless case. Such effective theories receive the name of {\it gauged supergravities}. For more details, see \cite{Grana:2005jc} and \cite{Samtleben:2008pe}.  \\

One interesting question, concerns the existence of spherically symmetric solutions in the context of gauged supergravities  \cite{Hristov:2010eu}-\cite{Hristov:2009uj}
 and in particular, the specific case in which a Kalb-Ramond (KR) three-form flux is extended in the effective four-dimensional space-time. The presence of KR fluxes implies the study of gravitational theories in curved backgrounds with torsion \cite{P}. Torsion appears as an antisymmetric tensorial piece in the Christoffel connection symbol in
Einstein-Cartan (EC) theory. The simplest level of torsion
theory can provide a classical background for quantum matter fields and
a relationship with spin \cite{V}-\cite{J}.  This is important since, an alternative way to study geometric aspects of general relativity at the microscopic level, is given by the EC theory \cite{Hehl}, which is characterized by the presence of a spin angular momentum in addition to the mass.  The main task to include torsion in space-time is to unify gravity with electromagnetism at a microscopic level,  where matter formations are performed by elementary particles.\\

Torsion can be incorporated as well into the classical solutions of Einstein equations, through a non-symmetric part of the KR metric, which in turns reduces to the effective antisymmetric two-index potential $B_{\mu \nu }$. The field strength of $B_{\mu\nu}$ has  in general,  a gauge invariant coupling
with the electromagnetic field,  leading to significant effects on many
cosmological and astrophysical phenomena \cite{4}-\cite{6}. \\

A complete understanding of torsion phenomena in the EC theory require
appropriate knowledge of the various solutions of the gravitational field
equations in space-times under various circumstances. Several cases have been studied in the last few years. For instance, in \cite{Sur}
SenGrupta and Sur obtained  the first known solution of Einstein equations in a space-time with torsion (effectively induced by  the presence of a four-dimensional KR field), corresponding to a static and spherically symmetric solution.  Later on, in \cite{6}, an approximate asymptotic solution was studied,  to analyze the gravitational lensing and the perihelion precession, and more over, it
was shown that this solution represents a wormhole model
for a real KR field (see also \cite{7}).\\

There are important progress in literature about the physical implications on the possible existence of torsion\footnote{We would like to thank Neil Russel for pointing out these references.}.  For instance, in \cite{15.12} it was reported tight bounds on many components of the torsion tensor (mixed-symmetry, trace and axial components) using the high-precision data from masers and torsion pendulums. See also \cite{15.13} for an improvement on the constraints on the axial components.\\

Motivated by  this, our goal in this paper is to explore
the possible existence of an evolving (time-dependent) spherically symmetric solution of the gravitational field equations (KR space-time) in presence of torsion\footnote{Although not explored in this work, such solutions could be important in order to realize the effect of torsion on electromagnetic waves coming from distant galactic sources \cite{cinco}.}. This means we shall start up with an effective four-dimensional action with a torsion component in the sense of Einstein-Cartan theory. The torsion component is given by the presence of a three-form flux which in turn is motivated from the string theory perspective.  In particular, following \cite{P} we consider an effective action which comes from considering a type IIB string compactification with Neveu-Schwarz (NS-NS) and Ramond-Ramond (RR) three-form fluxes with coordinates only on the extended space-time. Effectively those fluxes are interpreted as three-form fluxes which can be real or complex, according to the specific flux configuration selected in the string theory action.\\

Once we have specified the effective action,
we use group analysis as a method to find an analytical solution of non-linearized field Einstein-KR equation systems \cite{Ovsyannikov:1982}-\cite{15.1}. The group analysis lets us know if a differential system admits solutions with specific symmetries such as, for example, spherical symmetry. By this method, we obtain a set of
solutions which are invariant  with respect to an
admitted group of point transformations. Some important works about symmetry analysis,  can be seen in  \cite{15.1}-\cite{15.3}, and some applications
to Einstein equations are found in \cite{15.1} and \cite{15.4}.  Therefore, by means of this method, we find an analytical evolving and spherically symmetric solution of Einstein equations with torsion.\\

Our paper is organized as follows: In section II we focus on
the action of a purely gravitational field theory in a background with torsion and we compute the corresponding equations of motion. They consist on two set of equations which must be reduced; this is done in
section III, where  we obtain explicitly the final system equations for the
Einstein-KR model. Using group analysis (analysis of symmetries), in section IV we find an evolving spherically symmetric
solution. In section V,  we take some appropriate limits on our solutions and we compare them with previous static solutions reported in literature. Finally we show our conclusions and briefly comment on some possible physical implications.


\section{Field equations for gauge-invariant Einstein-Cartan-Kalb-Ramond
coupling}

The presence of fluxes in string compactification yields the construction of a variety of four-dimensional effective actions depending on the way the fluxes are turned on. Generically, for the fluxes living on the internal space is possible to stabilize the moduli related to the compactification process itself.\\

For fluxes with coordinates on the extended four-dimensional space-time after compactification, effective actions can describe a  curved space-time with torsion as shown in \cite{1}, where it was considered the presence of a Neveu-Schwarz - Neveu-Schwarz (NS-NS) three-form flux in a ten-dimensional type IIB superstring with a bosonic action given by
\begin{equation}
S=\frac{1}{2\kappa_{10}^2}\int d^{10}x\sqrt{-G}\left(R-\frac{\tau\bar\tau}{2Im~\tau}H_{MNP}H^{MNP}\right),
\end{equation}
in the Einstein-frame\footnote{Where we have used that the Einstein-frame and string-frame metrics are related by $G^E_{MN}=e^{-\phi/2}G^S_{MN}$.}, and where $\tau=C_0+ie^{-\phi}$ is the complex scalar formed by the dilaton $\phi$ and the Ramond-Ramond (RR) zero-form $C_0$.\\

After compactification\footnote{We follow the standard notation on which capital indices $M,N,P,..$ are used to denote ten-dimensional coordinates, while greek indices $\mu,\nu,\lambda,...$ are used to describe four-dimensional extended coordinates.} and
by considering a space-time threaded with a three-form flux $H_{\mu\nu\rho}$, this field induces a torsion component on the connection\footnote{As briefly commented in the introduction, the torsion component can be elucidated from a stringy point of view. By requiring the effective four-dimensional theory to have 4 supersymmetric generators, the presence of three-form fluxes contributes with an extra term in the ten-dimensional gravitino variation, establishing the presence of an effective torsion component on the four-dimensional connection. Different flux configurations lead to a variety of four-dimensional gauged supergravities.  By considering a simple configuration given by a single 3-form flux extended in four-dimensions, EC action should describe the bosonic part of  a specific gauged supergravity theory \cite{CRO}.}, yielding an effective $(3+1)-$ dimensional and gauge-invariant action, with an Einstein-Cartan-Kalb-Ramond (ECKR) coupling in presence of
external matter fields (we shall closely follow \cite{P}). The action is given by
\begin{eqnarray}
S =\frac{1}{\kappa}\int \sqrt{-g}\left( \tilde{R}\left( g,\mathcal{T}\right) -\frac{\kappa}{2}H_{\mu \nu \lambda }H^{\mu \nu \lambda }+\kappa\mathcal{T}^{\mu \nu \lambda }H_{\mu \nu
\lambda }\right)\,d^{4}x\,,
\label{SKRa}
\end{eqnarray}%
where $\tilde{R}\left( g,\mathcal{T}\right)$ is the scalar curvature of the
EC space-time and $\mathcal{T}^{\mu \nu \lambda }$ is the connection that contains
the torsion tensor.  Notice that  $\mathcal{T}^{\mu \nu \lambda }$ is the anti-symmetrization of the affine
connection in EC space-time. As usual,  $H_{\mu \nu \eta }=\partial _{\lbrack \mu }B_{\nu \eta ]}$ is
the strength of the KR field, $B_{\mu \nu }$ and $\kappa =16\pi G$ is the corresponding coupling constant.\\

As shown in \cite{P}, the curvature scalar $\tilde{R}\left( g,%
\mathcal{T}\right) $ of the EC space-time can be written in terms of the curvature of a purely
Riemmannian space-time $R(g)$ and the corresponding torsion tensor $\mathcal{T}_{\mu\nu\lambda}$ as%
\begin{equation}
\tilde{R}\left( g,\mathcal{T}\right) =R(g)+\kappa\mathcal{T}_{\mu \nu \lambda }%
\mathcal{T}^{\mu \nu \lambda },
\end{equation}%
where $\mathcal{T}_{\mu\nu\lambda}$ satisfies the constraint equation%
\begin{equation}
\mathcal{T}_{\mu \nu \lambda }=-\frac{1}{2}H_{\mu \nu \lambda }.
\label{Tmnl}
\end{equation}%
Therefore, it is followed that the KR field strength $H_3$ plays the role
of the spin angular momentum density, considered to be the source of
torsion \cite{1}. \\

In our case,  we start from the ten-dimensional action in the presence of both, NS-NS and RR three-form fluxes $H_{MNP}$ and $F_{MNP}$ respectively. The corresponding action is given by
 \begin{equation}
S=\frac{1}{2\kappa_{10}^2}\int d^{10}x\sqrt{-G}\left(R-\frac{\tau\bar\tau}{2Im~\tau}\mathcal{G}_{MNP}\bar{\mathcal{G}}^{MNP}\right),
\end{equation}
where $\mathcal{G}_{MNP}=\tau^{-1}F_{MNP}- H_{MNP}$ is a complex three-form. Notice that by turning off the RR field $F_3$, the remnant is a real field. Now, after compactification and by considering the complex three-form flux to have components on the extended four-dimensional space-time only,  a torsion term in the effective four-dimensional theory is induced. This  is  analogous to the term in the action given by Eq. (\ref{SKRa}), with ${H}_{\mu\nu\lambda}$ being substituted by $\cal{G}_{\mu\nu\lambda}$. The effective action reads
\begin{equation}
S =\frac{1}{\kappa}\int \sqrt{-g}\left( \tilde{R}\left( g,\mathcal{T}\right) -\frac{\kappa}{2}{\cal G}_{\mu \nu \lambda }\bar{\cal G}^{\mu \nu \lambda }+\kappa\mathcal{T}^{\mu \nu \lambda }{\cal G}_{\mu \nu
\lambda }\right)\,d^{4}x\,,
\label{SKR}
\end{equation}
which we shall also refer to as ECKR action. \\

Therefore, after making use of the constraint on the torsion element, the corresponding equations of motion are
\begin{eqnarray}
G_{\mu \nu } &\equiv &R_{\mu \nu }-\frac{1}{2}g_{\mu \nu }R=-\kappa T_{\mu
\nu },  \label{EE} \\
D_{\mu }\cal{G}^{\mu \nu \lambda } &\equiv &\frac{1}{\sqrt{\kappa }}\partial _{\mu
}\left( \sqrt{-g}\cal{G}^{\mu \nu \lambda }\right) =0,  \label{EM}
\end{eqnarray}%
where $T_{\mu \nu }$, a symmetric 2-tensor, plays the role of the
energy-momentum tensor\footnote{Notice that this tensor is directly related to the spin-density
tensor in the field equations for the ECKR space-time},  which reads

\begin{equation}
T_{\mu \nu }=\frac{3}{4}\left( 3g_{\nu \rho }{\cal G}_{\alpha \beta \mu }\bar{\cal G}^{\alpha
\beta \rho }-\frac{1}{2}g_{\mu \nu }{\cal G}_{\alpha \beta \gamma }\bar{\cal G}^{\alpha \beta
\gamma }\right) .  \label{TmnS}
\end{equation}

Following the notation used in \cite{Sur},  let us denote by $h_{1}$, $h_{2}$, $h_{3}$ and $h_{4}$ the covariant four independent complex components, ${\cal G}_{012}$,  ${\cal G}_{013}$,  ${\cal G}_{023}$ and  ${\cal G}_{123}$,  of the complex flux ${\cal G}_3$, and by $h^{1}$, $h^{2}$, $h^{3}$ and $h^{4}$ the corresponding complex-conjugated contravariant components. Notice as well that from Einstein equations, the curvature of the space-time is given by
\begin{equation}
R=-\kappa|{\cal G}_3|^2,
\label{torsioncurv}
\end{equation}
where $|{\cal G}_3|^2={\cal G}_{\mu\nu\lambda}\bar{\cal G}^{\mu\nu\lambda}$. Therefore we expect that a solution of the Einstein equations in the presence of torsion, describes a space-time with a negative curvature proportional to the flux, which in turn, is the source of torsion. \\

Recall that our goal is to find time-dependent exact solutions for the system of Eqs. (\ref{EM}). For that,  we shall consider the most general metric with an evolving spherical symmetry,
\begin{equation}
ds^{2}=\Omega ^{2}(t)\left[ e^{\nu }dt^{2}-e^{\lambda }dr^{2}-r^{2}(d\theta
^{2}+\sin ^{2}{\theta }d\phi ^{2})\right] \,,  \label{EWH}
\end{equation}%
where $\nu =\nu (t,r),$ $\lambda =\lambda (t,r)$ and $\Omega =\Omega (t)$ is
the conformal factor, which is finite and positive-definite over the
domain of $t$.  Therefore, from Eqs. (\ref{EE}) and (\ref{TmnS}) we obtain the following set of five differential
equations, corresponding to the non-zero components of Einstein equations:
\begin{eqnarray}
{\bar{\kappa}\Omega }^{2}\left(
h^{1}h_{1}+h^{2}h_{2}+h^{3}h_{3}-h^{4}h_{4}\right) &=&  {e}^{-\lambda }\left( -\frac{\lambda ^{\prime }}{r}+\frac{1}{r^{2}}\right)
{-}\frac{1}{r^{2}}-{e}^{-\nu }\left[ \dot{\lambda}\frac{\dot{\Omega}}{\Omega
}+3\left( \frac{\dot{\Omega}}{\Omega }\right) ^{2}\right] ,  \label{EE1} \\
 \notag \\
 {\bar{\kappa}\Omega }^{2}\left(
h^{1}h_{1}+h^{2}h_{2}-h^{3}h_{3}+h^{4}h_{4}\right) &=&  {e}^{-\lambda }\left( \frac{\nu ^{\prime }}{r}+\frac{1}{r^{2}}\right) {-}%
\frac{1}{r^{2}}{+e}^{-\nu }\left[ \left( \frac{\dot{\Omega}}{\Omega }\right)
^{2}-2\frac{\ddot{\Omega}}{\Omega }+\dot{\nu}\frac{\dot{\Omega}}{\Omega }%
\right] ,  \label{EE2}\\
\notag\\
{2\bar{\kappa}\Omega }^{2}\left(
h^{1}h_{1}-h^{2}h_{2}+h^{3}h_{3}+h^{4}h_{4}\right) &=&  {e}^{-\lambda }\left( \nu ^{\prime \prime }+\frac{\nu ^{\prime 2}}{2}-%
\frac{\nu ^{\prime }\lambda ^{\prime }}{2}+\frac{\nu ^{\prime }-\lambda
^{\prime }}{r}\right) {-e}^{-\nu }\left( \ddot{\lambda}+\frac{\dot{\lambda}%
^{2}}{2}-\frac{\dot{\nu}\dot{\lambda}}{2}\right)\label{EE3}\\
&&+{2e}^{-\nu }\left[ \left( \frac{\dot{\Omega}}{\Omega }\right) ^{2}-2\frac{%
\ddot{\Omega}}{\Omega }+\left( \dot{\nu}-\dot{\lambda}\right) \frac{\dot{%
\Omega}}{\Omega }\right] , \notag \\
&&  \notag\\
{2\bar{\kappa}\Omega }^{2}\left(-h^{1}h_{1}+h^{2}h_{2}+h^{3}h_{3}+h^{4}h_{4}\right) &=& {e}^{-\lambda }\left( \nu ^{\prime \prime }+\frac{\nu ^{\prime 2}}{2}-%
\frac{\nu ^{\prime }\lambda ^{\prime }}{2}+\frac{\nu ^{\prime }-\lambda
^{\prime }}{r}\right) {-e}^{-\nu }\left( \ddot{\lambda}+\frac{\dot{\lambda}%
^{2}}{2}-\frac{\dot{\nu}\dot{\lambda}}{2}\right)  \label{EE4} \\
&&+{2e}^{-\nu }\left[ \left( \frac{\dot{\Omega}}{\Omega }\right) ^{2}-2\frac{%
\ddot{\Omega}}{\Omega }+\left( \dot{\nu}-\dot{\lambda}\right) \frac{\dot{%
\Omega}}{\Omega }\right] ,  \notag \\
&&  \notag \\
2{\bar{\kappa}}\Omega ^{2}h_{3}h^{4}&=&e^{-\lambda }\left[ \frac{\dot{\lambda%
}}{r}+\nu ^{\prime }\frac{\dot{\Omega}}{\Omega }\right] ,  \label{EE5}
\end{eqnarray}%
where an over-dot denotes a derivative with respect to the time coordinate $t$,  a prime  denotes a derivative with respect to $r$ and ${\bar{\kappa}=}\frac{9}{4}\kappa $.  On the other hand, the vanishing components of Einstein equations, provide with a set of constraints on the $h$ fluxes, given by
\begin{equation}
h^{4}h_{2}=h^{4}h_{1}=h^{3}h_{2}=h^{2}h_{1}=h^{1}h_{3}=0.  \label{EE6}
\end{equation}
Besides the above differential equation, there is another set of equations corresponding to the dynamics of the field $H_3$ given by Eq. (\ref{EM}). These six independent equations read
\begin{eqnarray}
0 &=&h_{,2}^{1}+h^{1}\cot \theta +h_{,3}^{2},  \label{Eh1} \\
0 &=&h_{,1}^{1}+\left( \frac{\nu ^{\prime }+\lambda ^{\prime }}{2}+\frac{2}{r%
}\right) h^{1}-h_{,3}^{3},  \label{Eh2} \\
0 &=&h_{,1}^{2}+\left( \frac{\nu ^{\prime }+\lambda ^{\prime }}{2}+\frac{2}{r%
}\right) h^{2}+h_{,2}^{3}+h^{3}\cot \theta ,  \label{Eh3} \\
0 &=&h_{,0}^{1}+\left( \frac{\dot{\nu}+\dot{\lambda}}{2}+4\frac{\dot{\Omega}%
}{\Omega }\right) h^{1}+h_{,3}^{4},  \label{Eh4} \\
0 &=&h_{,0}^{3}+\left( \frac{\dot{\nu}+\dot{\lambda}}{2}+4\frac{\dot{\Omega}%
}{\Omega }\right) h^{3} +h_{,1}^{4}+\left( \frac{\nu ^{\prime }+\lambda ^{\prime }}{2}+\frac{2}{r}%
\right) h^{4},  \label{Eh5} \\
0 &=&h_{,0}^{2}+\left( \frac{\dot{\nu}+\dot{\lambda}}{2}+4\frac{\dot{\Omega}%
}{\Omega }\right) h^{2}-h_{,2}^{4}-h^{4}\cot \theta ,  \label{Eh6}
\end{eqnarray}%
where the subindex $i$ runs over $(0,1,2,3)$ and  denotes a derivative with respect to the corresponding $(t,r,\theta ,\phi )$ variables.\\

The system of differential equations is reduced by Eqs. (\ref{EE3})-(\ref{EE4}), which implies a vanishing values for the fluxes $h_1$ and $h_2$.  Notice that these flux values satisfy the constraint imposed by the zero components of Einstein equations given by Eq. (\ref{EE6}). More reductions can be performed by considering Eq. (\ref{Eh2}) and Eq. (\ref{Eh4}). We select an axial symmetry on the system, such that
\begin{equation}
h_{,3}^{3}=h_{,3}^{4}=0.
\end{equation}
It is important to stress out, that this condition allows the parameter $\lambda$ and $\omega$ to be time-dependent, leading us to the construction of a corresponding time-dependent solution.\\

The last conditions, reduces the system of Eqs. (\ref{Eh1})-(\ref{Eh6}) to
three differential equations, given by
\begin{eqnarray}
0 &=&h_{,2}^{3}+h^{3}\cot \theta,  \label{SE1} \\
0 &=&h_{,0}^{3}+\left( \frac{\dot{\nu}+\dot{\lambda}}{2}+4\frac{\dot{\Omega}%
}{\Omega }\right) h^{3}  +h_{,1}^{4}+\left( \frac{\nu ^{\prime }+\lambda ^{\prime }}{2}+\frac{2}{r}%
\right) h^{4},  \label{SE2}  \\
0 &=&-h_{,2}^{4}-h^{4}\cot \theta.  \label{SE3}
\end{eqnarray}
The corresponding solutions for $h_3$ and $h_4$ read
\begin{eqnarray}
h^{3} &=&\frac{h_{3}}{\Omega ^{6}r^{4}\sin^{2}\theta e^{\nu }}=\frac{%
u_{3}(t,r)}{\sin \theta},  \label{I1} \\
h^{4} &=&\frac{-h_{4}}{\Omega ^{6}r^{4}\sin^{2}\theta e^{\lambda }}=\frac{%
u_{4}(t,r)}{\sin \theta},  \label{I2}
\end{eqnarray}
where we have expressed them in terms of the functions $u_3$ and $u_4$, both depending on the parameters $t$ and $r$.\\

Now, by substituting Eq. (\ref{I1}) and Eq. (\ref{I2}) into Eq. (\ref{SE2}%
) and into Eqs. (\ref{EE1})--(\ref{EE5}) we obtain the following system of differential equations,
\begin{eqnarray}
\dot{u}_{3}+u_{4}^{\prime }&=&-\left( \frac{\dot{\nu}+\dot{\lambda}}{2}+4%
\dot{w}\right) u_{3}  -\left( \frac{\nu ^{\prime }+\lambda ^{\prime }}{2}+\frac{2}{r}\right)
u_{4},  \label{SSE3} \\
\bar{\kappa}e^{8w}r^{4}\left( u_{3}^{2}e^{\nu }+u_{4}^{2}e^{\lambda
}\right) &=&{e}^{-\lambda }\left( -\frac{\lambda ^{\prime }}{r}+\frac{1}{r^{2}}%
\right)  -\frac{1}{r^{2}}{-e}^{-\nu }\left[ \dot{\lambda}\dot{w}+3\dot{w}^{2}\right]
,  \label{SE4} \\
-\bar{\kappa}e^{8w}r^{4}\left( u_{3}^{2}e^{\nu }+u_{4}^{2}e^{\lambda
}\right)& =&{e}^{-\lambda }\left( \frac{\nu ^{\prime }}{r}+\frac{1}{r^{2}}%
\right)  -\frac{1}{r^{2}}{+e}^{-\nu }\left[ \dot{\nu}\dot{w}-2\ddot{w}-\dot{w}^{2}%
\right] ,  \label{SE5} \\
2\bar{\kappa}e^{8w}r^{4}\left( u_{3}^{2}e^{\nu }-u_{4}^{2}e^{\lambda
}\right) &=&{-e}^{-\nu }\left( \ddot{\lambda}+\frac{\dot{\lambda}^{2}}{2}-%
\frac{\dot{\nu}\dot{\lambda}}{2}\right) {+e}^{-\lambda }\left( \nu ^{\prime \prime }+\frac{\nu ^{\prime 2}}{2}-%
\frac{\nu ^{\prime }\lambda ^{\prime }}{2}+\frac{\nu ^{\prime }-\lambda
^{\prime }}{r}\right)  \notag\\
&&+{2e}^{-\nu }\left[ (\dot{\nu}-\dot{\lambda})\dot{w}-2\ddot{w}-\dot{w}^{2}%
\right] ,  \label{SE6}\\
2\bar{\kappa}e^{8w+\nu }r^{4}u_{3}u_{4}&=&e^{-\lambda }\left[ \frac{\dot{%
\lambda}}{r}+\nu ^{\prime }\dot{w}\right] ,  \label{SE8}
\end{eqnarray}%
where we have taken $\Omega (t)=e^{w(t)}$. We are now ready to apply
the group analysis method in order to obtain an evolving exact solution, since our system has been reduced till we have the same number of equations and variables. This shall be performed in the next section.


\section{Kalb-Ramond solution using group analysis}

\subsection{Symmetry analysis and invariant solution forms}

To begin with, let us describe briefly the group analysis method \cite{Ovsyannikov:1982} we shall use to find invariant solutions of the system of differential equations obtained in the last section. In pedestrian words, the method consists on three steps\footnote{For a more extensive description of this method, see \cite{15} and \cite{15.3}.}. First, it is necessary to find
out the Lie algebra of generators of the point-transformations group. This group is constructed according to a particular  system of differential equations. In our case, the system of Eqs. (\ref{SSE3})-(\ref{SE8}) has a specific group which we shall consider.
Second, the sub-algebras, with
respect to the inner automorphisms group, are classified. This allows to construct some forms related to the
invariant solutions. Finally, substituting the invariant-forms,  we construct a system from which we obtain the
so-called factor-system, that in our case, consists of an ordinary
differential equations system. By solving the factor-system, we construct an  analytic invariant solution.\\

Therefore, we start by describing the infinitesimal generators admitted by the system. Since we have a system of five equations, depending on two variables, in the most general form, the infinitesimal generators are given by
\begin{equation}
X=\xi^{i}(x^1,x^2)\frac{\partial}{\partial x^{i}}+\eta^{k}(u^1,\dots,u^5)%
\frac{\partial}{\partial u^{k}},
\end{equation}
where $x^1=t$, $x^2=r$, $(u^1,...,u^5)$ = $(u_3,u_4,\lambda,\nu,w)$, $i =
1,2$ and $k=1,\dots,5$.\\

At this point, it is necessary to prolongate the group operator by extending the infinitesimal generators. In case of a
second order system, the corresponding prolongated-generators are given by
\begin{eqnarray}
\underset{2}X&=&X+\xi _{i}^{\alpha }\frac{\partial }{\partial p_{i}^{\alpha }}+\xi
_{ij}^{\alpha }\frac{\partial }{\partial p_{ij}^{\alpha }}, \\
 \xi _{i}^{\alpha }&=&D_{i}(\eta ^{\alpha })-p_{\beta }^{\alpha }D_{i}(\xi
^{\beta }), \\
\xi _{ij}^{\alpha }&=&D_{j}(\xi _{i}^{\alpha })-p_{i\beta }^{\alpha
}D_{j}(\xi ^{\beta }),
\end{eqnarray}
where $D_{i}=\frac{\partial }{\partial x^{i}}+p_{i}^{\alpha }\frac{\partial }{%
\partial u^{\alpha }}+p_{ij}^{\alpha }\frac{\partial }{\partial
p_{j}^{\alpha }}+\dots$ is operator of total derivative, $p_{i}^{\alpha }=\dfrac{\partial u^{\alpha }}{\partial x^{i}}$, $%
p_{ij}^{\alpha }=\dfrac{\partial ^{2}u^{\alpha }}{\partial x^{i}\partial
x^{j}}$ and $\alpha =1,\dots ,5$, $\beta =1,2$, $j=1,2$.\\

By applying  the prolongate group operator to every equation in our original system,
we are able to solve the
overdetermined system of differential equations with respect to coefficients
$\xi ^{i}(x^{1},x^{2})$ and $\eta ^{k}(u^{1},\dots ,u^{5})$. Schematically, this is given by the equation
$\left. \underset{2}X(F)\right\vert _{F=0}=0$, where $F=0$ denotes the initial differential equations system. This allows to
find out the Lie algebra satisfied by the
infinitesimal generators admitted by the system of Eqs. (\ref{SSE3})-(\ref%
{SE8}). This Lie algebra, denoted $L_{3}$,  has the following basis,
\begin{eqnarray}
X_{1} &=&t\frac{\partial }{\partial t}+r\frac{\partial }{\partial r}-3u_{3}%
\frac{\partial }{\partial u_{3}}-3u_{4}\frac{\partial }{\partial u_{4}}, \\
X_{2} &=&\frac{\partial }{\partial t}, \\
X_{3} &=&\frac{\partial }{\partial w}-4u_{3}\frac{\partial }{\partial u_{3}}%
-4u_{4}\frac{\partial }{\partial u_{4}},
\end{eqnarray}
which satisfy the commutator relation $\lbrack X_{1},X_{2}]=-X_{2}$.\\

Besides, in relation to the group of the inner automorphisms, there are  nonequivalent
sub-algebras $L_{3}$, which according to the classification in
\cite{16}, are given by the generators
\begin{eqnarray}
&&X_{2}+\gamma X_{3},  \label{A1} \\
&&X_{1}+\gamma X_{3},  \label{A2} \\
&&X_{3},  \label{A3}
\end{eqnarray}%
where $\gamma $ is a real arbitrary parameter. The sub-algebras generated by Eq. (\ref{A1}) and Eq. (\ref{A2}) provide the invariant solution forms. Notice that for
the sub-algebra corresponding to the generator Eq. (\ref{A3}),  there is not an associated invariant solution because it does not
contain any independent variable.\\

In order to get the invariant solution form, corresponding to the sub-algebra Eq. (\ref{A1}), it is necessary to solve the first
order differential equations, given by
\begin{equation}
\left( X_{2}+\gamma X_{3}\right) \Phi =0,
\end{equation}
and find all invariants $\Phi$. The above differential equation implies that
\begin{equation}
dt=-\frac{du_{3}}{4\gamma t}=-\frac{du_{4}}{4\gamma t}=-\frac{dw}{\gamma }.
\end{equation}
Therefore,  the set of invariants is given by
\begin{eqnarray}
\Phi _{0} =r,\ \Phi _{1}=\lambda ,\ \Phi _{2}=\nu , \Phi_{3}=u_{3}e^{4\gamma t}, \Phi _{4} =u_{4}e^{4\gamma t},\ \Phi _{5}=w-\gamma t.
\end{eqnarray}
Now, taking $\Phi _{i}=\Phi _{i}(\Phi _{0})$, where $i=1,\dots ,5,$ we obtain the first form of an invariant solution
\begin{eqnarray}
&&u_{3}=\Phi _{3}(r)e^{-4\gamma t},\ u_{4}=\Phi _{4}(r)e^{-4\gamma t}, w=\Phi _{5}(r)+\gamma t,\ \lambda =\Phi _{1}(r),\ \nu =\Phi _{2}(r).
\end{eqnarray}
Now, since $\dfrac{\partial w}{\partial r}=0$,  it follows that $\Phi _{5}(r)=m=\text{const}$. Henceforth, the first form of the
invariant solutions is given by
\begin{eqnarray}
&&u_{3}(r,t)=\Phi _{3}(r)e^{-4\gamma t}, \quad u_{4}(r,t)=\Phi
_{4}(r)e^{-4\gamma t},  \quad w(t)=m+\gamma t,\quad \lambda (r)=\Phi _{1}(r),\quad \nu (r)=\Phi _{2}(r),
\label{SS1}
\end{eqnarray}
where $m$ is a real parameter and $\Phi _{i}$ ($i=1,\dots ,4$) are undetermined functions. It is straightforward to notice  here that this invariant solution
corresponds to a stationary system, since  both $\lambda $ and $\nu$,  depend only on
$r$. In other words, the evolving
spherically symmetric solution is obtained from the first form of invariant solution.\\

By a similar procedure, the second invariant solution form corresponding to the sub-algebra  (%
\ref{A2}), reads
\begin{eqnarray}
u_{3}(z,t)=\Phi _{3}(z)(tr)^{-(3+4\gamma )/2}, \ \lambda (z)=\Phi _{1}(z), \ u_{4}(z,t)=\Phi _{4}(z)(tr)^{-(3+4\gamma )/2},\notag\\
w(t)=m+\ln t^{\gamma },
  \  \nu (z)=\Phi _{2}(z),\  z=\dfrac{t}{r}.
\label{SS2}
\end{eqnarray}

We have found two forms of invariant solutions, which allow us to construct two different kind of solutions of our original system of differential equations. We shall find an evolving spherically symmetric solution and a dynamical one. These are the next and final steps.

\subsection{Evolving spherically symmetric solution}

Let us consider both
$\lambda$ and $\nu$ to be time-invariant. Our aim is to find an evolving non-vacuum solution in a background with torsion.
Making use of the set of identities solutions described in Eqs. (\ref{SS1}),
(\ref{I1}) and (\ref{I2}), we proceed to substitute them into the system of Eqs. (%
\ref{SSE3})--(\ref{SE8}), in order to obtain the factor-system of the
ordinary differential equations for the undetermined functions $\Phi _{i}(r)$, which is given by
\begin{eqnarray}
0&=&2\Phi _{4}^{\prime }+\left( \frac{4}{r}-\frac{\psi _{2}^{\prime }}{\psi
_{2}}-\frac{\psi _{1}^{\prime }}{\psi _{1}}\right) \Phi _{4},  \label{EN1} \\
0&=&-k_{1}r^{6}\left( \frac{\Phi _{3}^{2}}{\psi _{2}}+\frac{\Phi _{4}^{2}}{%
\psi _{1}}\right) +r\psi _{1}^{\prime }+\psi _{1}   -3\gamma ^{2}r^{2}\psi _{2}-1,  \label{EN2}\\
0&=&k_{1}r^{6}\left( \frac{\Phi _{3}^{2}}{\psi _{2}}+\frac{\Phi _{4}^{2}}{\psi _{1}}\right) -r\frac{\psi _{1}}{\psi _{2}}\psi _{2}^{\prime }+\psi _{1}
-\gamma ^{2}r^{2}\psi _{2}-1,  \label{EN3} \\
0&=&2k_{1}r^{5}\left( \frac{\Phi _{4}^{2}\psi _{2}}{\psi _{1}}-\Phi
_{3}^{2}\right) -r\psi _{1}\psi _{2}^{\prime \prime }-\frac{1}{2}r\psi
_{2}^{\prime }\psi _{1}^{\prime }
+\frac{3r\psi _{1}}{2\psi _{2}}\left( \psi _{2}^{\prime }\right) ^{2}+\psi
_{2}\psi _{1}^{\prime }-\psi _{1}\psi _{2}^{\prime }-2\gamma ^{2}r\psi
_{2}^{2},   \label{EN4}  \\
0&=&2k_{1}r^{4}\frac{\Phi _{3}\Phi _{4}}{\psi _{1}\psi _{2}}+\gamma \frac{%
\psi _{2}^{\prime }}{\psi _{2}},  \label{EN5}
\end{eqnarray}
where $k_{1}=\bar{\kappa}e^{8m}>0$, $\psi _{1}=e^{-\Phi _{1}}>0$ and $\psi
_{2}=e^{-\Phi _{2}}>0$.\\

In order to solve the system of Eqs. (\ref{EN1})-(\ref{EN5}), first we analyze their
compatibility conditions. This analysis gives us some long
terms, so we limit ourselves to explain only the process to reach the important additional
information. Assuming $\Phi _{4}\neq 0$ (to avoid singularities into the equations), we solve Eq. (\ref{EN5}) for $\Phi _{3}$
and substitute this solution into the system of Eqs. (\ref{EN1})-(\ref%
{EN4}). Under this substitution,  Eq. (\ref{EN2}) becomes a quadratic equation with
respect to $\Phi _{4}$. Solving for $\Phi _{4}$, we substitute the corresponding solution into Eqs. (\ref%
{EN1}), (\ref{EN3}) and (\ref{EN4}), such that we obtain a system of equations for $\psi _{1}$ and $ \psi _{2}$.\\

The obtained system reduces Eq. (\ref{EN3}), such that we get that $\psi _{2}^{\prime
}=0 $, which means that $\psi _{2}=\xi $, with $\xi $ being a real constant. The
same result can be obtained from Eq. (\ref{EN1}). If we substitute $\psi
_{2}=\xi $ into Eq. (\ref{EN4}), we obtain the following differential
equation for $\psi _{1}$,
\begin{eqnarray}
0 =\frac{1}{2} +\frac{1}{8}r^{2}\left( \psi _{1}^{\prime }\right) ^{2}+\left( \psi
_{1}-1-\frac{5}{2}\gamma ^{2}r^{2}\xi \right) r\psi _{1}^{\prime } +\left( \frac{1}{2}\psi _{1}-1-3\gamma ^{2}r^{2}\xi \right) \psi
_{1}+\left( 4\gamma ^{2}r^{2}\xi +3\right) \gamma ^{2}r^{2}\xi .  \label{EN33}
\end{eqnarray}
This last equation has the following two solutions
\begin{eqnarray}
\psi _{1,a} &=&\xi \gamma ^{2}r^{2}+1+\frac{c_{1}}{r^{2}},  \label{SSS1} \\
\psi _{1,b} &=&\xi \gamma ^{2}r^{2}+1+\frac{c_{1}}{r^{\frac{2}{3}}},
\label{SSS2}
\end{eqnarray}
where $c_{1}$ is an arbitrary constant. Using $\psi _{2}=\xi $ and Eq. (\ref%
{SSS1}), we can see from Eq. (\ref{EN5}) that there are two kind of solutions: $\Phi
_{3}=0$ or $\Phi _{4}=0$. If we choose $\Phi _{3}=0$ and Eq. (\ref{SSS1}) or
Eq. (\ref{SSS2}) to solve the system, then necessarily $\Phi _{4}=0$. Now,
if we take Eq. (\ref{SSS2}) and $\Phi _{4}=0$ the solution must be $\Phi
_{3}=0$ because we need $c_{1}=0$. Therefore, there is a unique solution in which a non-trivial torsion source can be described. This is achieved by taking $\Phi_4=0$ and by selecting Eq. (\ref{SSS1}).\\

Choosing $\Phi_{4}=0$ and substituting it into Eqs. (\ref{EN1})-(\ref{EN5}), we obtain from Eq. (\ref{EN4}), an algebraic equation for $\Phi_{3}$,
\begin{eqnarray*}
0 &=&k_{1}r^{12}\Phi _{3}^{2}\xi \gamma ^{2}+k_{1}r^{10}\Phi
_{3}^{2}+k_{1}r^{8}\Phi _{3}^{2}c_{1} +\xi c_{1}r^{2}+\xi c_{1}^{2}+\xi ^{2}\gamma ^{2}c_{1}r^{4},
\end{eqnarray*}
which solution reads,
\begin{equation}
\Phi _{3}=\pm \sqrt{\frac{\xi }{k_{1}}}\frac{c_{2}}{r^{4}},  \label{SSS3}
\end{equation}
where $c_{2}=\sqrt{-c_{1}}$ and $c_{1}$ is allowed to be positive or negative. Important to notice that, if $c_1$ is negative, $c_2$ is a real positive number and therefore, the corresponding flux $h^3$ is also real. In this case, we have considered implicitly that only NS-NS three-form fluxes were considered from a ten-dimensional string compactification. On the other hand, a positive $c_1$ imply the presence of complex three-form fluxes, from which we are considering string compactifications in the presence of both, RR and NS-NS three-form fluxes.\\

Finally, the complete set of solutions,  using Eqs. (\ref{SS1}), (\ref{SSS1}), (\ref{SSS3}), $\Phi _{4}=0$ and $\psi _{2}=\xi $ are given by
\begin{eqnarray}
u_{3}(r,t)&=&\frac{2 c_{2}}{3r^{4}}\sqrt{\frac{\xi}{\kappa}}e^{-4\left( m+\gamma
t\right) },\label{r}\\
\quad u_{4}(r,t)&=&0,  \\
e^{\lambda (r)}&=&\left(1+\dfrac{c_1}{r^{2}}+\xi\gamma
^{2}r^{2}\right)^{-1}, \notag\\
\quad e^{\nu (r)}&=&\xi ^{-1},  \\
\Omega (t)&=&e^{m+\gamma t}. \label{r4}
\end{eqnarray}

On the other hand (and notice that in general $u_3(r,t)$ is a complex function), from Eq. (\ref{I1}) we have the relation
\begin{equation}
h^{3}h_{3}=|u_{3}(r,t)|^2\Omega ^{6}r^{4}e^{\nu },
\label{met1}
\end{equation}
which reduces, once we considered the set of functions given by Eqs. (\ref{r})-(\ref{r4}), to
\begin{equation}
h^{3}h_{3}=\frac{|c_{2}|^{2}}{\bar{\kappa}r^{4}}e^{-2\left( m+\gamma t\right)
}. \label{T22}
\end{equation}

Since we obtained that $h_4=0$, from Eq. (\ref{r}) we see immediately  that a non-trivial solution requires that
$c_{2}$ and $\xi$ are both different from zero. Recall that torsion components in the connection are related to the presence of a flux ${\cal G}_3$. In our present case, the only component which survives is $h_3={\cal G}_{023}$. The squared flux, as seen from Eq.(\ref{T22}), goes as $r^{-4}$ at constant time, decaying to zero at infinity, faster than an ordinary electromagnetic field. Therefore, the corrections expected by torsion at large distances, through the presence of a 3-form flux, are smaller than those given by a Maxwell classical field. Notice also that,  as time runs to infinity, the solution decreases if and only if  $\gamma >0$ and $m>0$. Therefore, we shall consider this to be the case. In such conditions, we have found an evolving spherically symmetric solution of the Einstein equations in presence of torsion. Finally, let us mention that, as far as we know, there are no reports in literature about evolving exact solutions under these conditions. \\

\subsection{Dynamic solution, second invariant solution form}

In this section, we will analyze the ECKR dynamic solutions, which represents an
evolutionary system in time. We shall consider both, $\nu $ and $\lambda $, to be functions of $t$
and $r$. Using the second invariant solution form given by Eq. (\ref{SS2}) together with Eqs.
(\ref{SSE3})-(\ref{SE8}), we obtain the following factor-system of ordinary
differential equations,
\begin{eqnarray}
0 &=&z\left( \frac{\psi _{1}^{\prime }}{\psi _{1}}+\frac{\psi _{2}^{\prime }%
}{\psi _{2}}\right) \left( z\Phi _{4}-\Phi _{3}\right) -2z^{2}\Phi
_{4}^{\prime }+2z\Phi _{3}^{\prime }   +\left( 4\gamma -3\right) \Phi _{3}+\left( 1-4\gamma \right) z\Phi _{4}, \label{EM1} \\
0 &=&\left( \frac{\psi _{2}}{\psi _{1}}\gamma -z^{2}\right) z\psi
_{1}^{\prime }-\left( \frac{\Phi _{3}^{2}}{\psi _{2}}+\frac{\Phi _{4}^{2}}{%
\psi _{1}}\right) z^{4\gamma -1}k_{1} +\psi _{1}z^{2}-3\psi _{2}\gamma ^{2}-z^{2},  \label{EM2}  \\
0 &=&\left( \frac{\psi _{1}}{\psi _{2}}z^{2}-\gamma \right) z\psi
_{2}^{\prime }+\left( \frac{\Phi _{3}^{2}}{\psi _{2}}+\frac{\Phi _{4}^{2}}{%
\psi _{1}}\right) z^{4\gamma -1}k_{1}   -\psi _{2}\gamma ^{2}+\psi _{1}z^{2}+2\psi _{2}\gamma -z^{2}, \label{EM3} \\
0 &=&\frac{3\psi _{1}^{2}}{2\psi _{2}}z^{4}\left( \psi _{2}^{\prime }\right)
^{2}-\frac{3\psi _{2}^{2}}{2\psi _{1}}z^{2}\left( \psi _{1}^{\prime }\right)
^{2}-\psi _{1}^{2}z^{4}\psi _{2}^{\prime \prime }  +\psi _{2}^{2}z^{2}\psi _{1}^{\prime \prime }+\left( -z^{2}\psi _{1}\psi
_{2}+2\psi _{2}^{2}\gamma \right) z\psi _{1}^{\prime } +4\gamma \psi
_{2}^{2}\psi _{1}  \label{EM4} \\
&&+ \frac{z^2}{2}\left( \psi _{2} - \psi _{1}z^{2}\right) \psi
_{1}^{\prime}\psi_{2}^{\prime }-\left( \psi _{1}^{2}z^{2}+2\gamma \psi
_{2}\psi _{1}\right) z\psi _{2}^{\prime }  +\left( \Phi _{4}^{2}\psi _{2}-\Phi _{3}^{2}\psi _{1}\right) 2z^{4\gamma
-1}k_{1}-2\gamma ^{2}\psi _{2}^{2}\psi _{1},  \notag \\
0 &=&\psi _{1}\psi _{2}^{\prime }\gamma - \psi _{2}\psi _{1}^{\prime }
-2z^{4\gamma -3}k_{1}\Phi _{3}\Phi _{4},  \label{EM5}
\end{eqnarray}%
where a prime denotes a derivative with respect to $z$. In this section, we use the same
definitions for $k_{1}, \psi _{1}$ and $\psi _{2}$ as in the last section,
but here $\psi _{1}$, $\psi _{2}$ also depend on $z$.\\

It is easy to see that if $\Phi _{3}=z\Phi _{4}$, then Eq. (\ref{EM1}) is
fulfilled. Now from Eq. (\ref{EM5}) we have that
\begin{eqnarray}
&&\Phi _{4}^{2}=\frac{\psi _{1}\psi _{2}^{\prime }\gamma -\psi _{2}\psi
_{1}^{\prime }}{2k_{1}}z^{2-4\gamma },  \notag \\
&&\Phi _{3}^{2}=\frac{\psi _{1}\psi _{2}^{\prime }\gamma -\psi _{2}\psi
_{1}^{\prime }}{2k_{1}}z^{4-4\gamma }.  \label{fis}
\end{eqnarray}
Substituting Eqs. (\ref{fis}) into Eq. (\ref{EM2}) and Eq. (\ref{EM3}), and considering that $\gamma \neq -1$, we obtain
two equations for $\psi _{1}^{\prime }$ and $\psi _{2}^{\prime
}$,
\begin{eqnarray}
&&\psi _{1}^{\prime }=2\frac{\psi _{1}}{z}\frac{z^{4}\psi _{1}^{2}-z^{4}\psi
_{1}+\gamma ^{2}\psi _{2}^{2}-2\gamma ^{2}z^{2}\psi _{1}\psi _{2}}{%
(z^{2}\psi _{1}-\psi _{2})(z^{2}\psi _{1}-\gamma \psi _{2})},  \notag \\
&&\psi _{2}^{\prime }=-2\frac{\psi _{2}^{2}}{z}\frac{z^{2}(1-\psi
_{1}+\gamma \psi _{1})+\gamma (\gamma -1)\psi _{2}}{(z^{2}\psi _{1}-\psi
_{2})(z^{2}\psi _{1}-\gamma \psi _{2})}.  \label{deriv}
\end{eqnarray}
We can see that Eq. (\ref{EM4}) is satisfied identically, taking into
account Eq. (\ref{fis}) and Eq. (\ref{deriv}). In other words, any solution
of system Eqs. (\ref{deriv}) gives us a solution of the factor-system.\\

If we choose the special case in which $\gamma =-1$, hence Eq. (\ref{EM2}) is
equal to Eq. (\ref{EM3}). This lets us obtain one differential equation for the
two unknown functions $\psi _{1}$, $\psi _{2}$, which reads
\begin{equation}
\frac{\psi _{1}^{\prime }}{\psi _{1}}-\frac{\psi _{2}^{\prime }}{\psi _{2}}=%
\frac{2}{z}\frac{z^{2}\psi _{1}-3\psi _{2}-z^{2}}{\psi _{1}z^{2}+\psi _{2}}.
\label{uno}
\end{equation}

Making the same process to get (\ref{deriv}), we find that equation (\ref{EM4}) did not vanish as parallel to the previous case, so the resulting equation is,
\begin{eqnarray}
&&\frac{3z^2}{2}\left[ \frac{\psi_1^2z^2}{\psi_2}(\psi_2^\prime)^2-\frac{\psi_2^2}{\psi_1}(\psi_1^\prime)^2 \right]+ z^2(\psi_2^2\psi_1^{\prime\prime}-z^2\psi_1^2\psi_2^{\prime\prime})+\frac{z^2}{2}(\psi_2-z^2\psi_1)\psi_2^\prime\psi_2^\prime  \notag \\
&& -z\psi_2(3\psi_2\psi_1^\prime-\psi_1\psi_2^\prime)-6\psi_2^2\psi_1=0 \label{dos},
\end{eqnarray}
we can conclude for this special case that the factor system also reduces a couple of equations for two unknown variables, nevertheless the complexity of those equations in the general and special case give us the possibility to get a dynamical solution analysis for a future work.

\subsection{Exact evolving spherically symmetric solution for the metric with
torsion}

In this section we show the explicit metric form of the ECKR space-time for some
special cases. First, we substitute the solutions (corresponding to our first invariant solutions) given by Eqs. (\ref{r})-(\ref{r4}) into the metric shown in Eq. (\ref{EWH}). As a result, we obtain the metric for the ECKR space-time, which is given by\\
\begin{equation}
ds^{2}=e^{2\left( m+\gamma t\right) }\left[ \frac{dt^{2}}{\xi}-\frac{%
dr^{2}}{1+\frac{c_1}{r^{2}}+\xi\gamma ^{2}r^{2}}-r^{2}(d\theta
^{2}+\sin ^{2}{\theta }d\phi ^{2})\right], \,  \label{met}
\end{equation}
and the associated scalar curvature reads\\
\begin{eqnarray}
R=-\frac{2}{e^{2(m+\gamma t)}}\frac{c_1}{r^4}.
\label{sc}
\end{eqnarray}
We see that for large distances and time running to large values, the scalar curvature reaches a zero constant value. Therefore, this space-time evolves into a an asymptotically flat space-time. Notice that the parameter $\xi$ does not play a role in the curvature values, which can be positive or negative according to the kind of fluxes, real or complex, giving rise to torsion. From the string theory perspective, this corresponds to turning on  only NS-NS fluxes or both, RR and NS-NS ones.\\

It is important to point out that the curvature of this solution behaves as  expected from Einstein equations as shown in Eq. (\ref{torsioncurv}), i.e., that the curvature is produced by the presence of the torsion component which in turn is given by the presence of the flux $h_3$. Therefore, in the absence of the 3-form flux $h_3$ the metric solution would describe a flat space-time. The curvature must be proportional to $|h_3|^2$ which by Eqs.(\ref{torsioncurv}) and (\ref{T22}), is given by Eq.(\ref{sc}).\\

\begin{figure}[t]
\begin{center}
\centering \epsfysize=8cm \leavevmode
\epsfbox{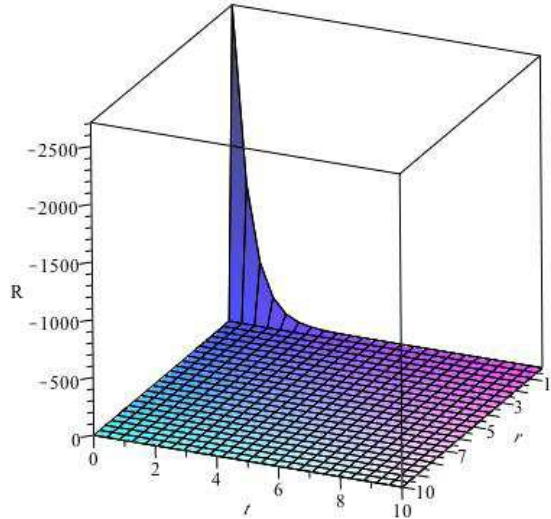}
\end{center}
\caption[hola]{\small \it Scalar curvature as a function of $r$ and $t$, with $\xi=\gamma=m=1$, and $c_1=1$.}
\label{small}
\end{figure}

Notice as well that as time runs to infinity, the curvature goes to zero even around the singularity at $r=0$. The corresponding space-time described by this solutions is highly curved around the singularity at $r=0$. However, as times runs, the curvature decreases and the space-time evolves into a soft universe with a null curvature for $r\neq 0$ (see Fig.\ref{small}). \\

Let us consider some specific values. First, notice that in the absence of flux, i.e., for $c_1=\gamma=m=0$,  a flat space-time is recovered.\\

\begin{enumerate}
\item
For non-trivial flux, the general solution depends on four parameters $(m, \gamma, \xi, c_1)$. One interesting case for the branch of solutions with $m=0$, involves a static configuration,  with
$\gamma =0$ and  $\xi
=1 $, and a negative $c_1$, in which the metric Eq. (\ref{met}) reduces to the wormhole solution found by
Sen Gupta et.al. in \cite{6}, where the horizon is found  at $r=c_2$.
Notice as well that the torsion term given by Eq. (\ref{T22}) reduces in this case to
\begin{equation}
h^{3}h_{3}=\frac{c_{2}^2}{\bar{\kappa} r^4},
\end{equation}
which represents an static flux-configuration in the presence of a real field axion (Hodge dual of the three-form flux). In this case, only NS-NS fluxes are taken into account in the ten-dimensional theory. \\

On the other hand,  this also suggests us that our exact solution Eq. (\ref{met}) with $\gamma\neq 0$ and $m=0$, should represent an evolving wormhole, although a more detailed analysis is required here\footnote{The presence of a three-form flux in the effective four-dimensional space-time, has been studied as well from the point of view of string compactifications.  A relation with extremal supersymmetric black holes is shown in \cite{LoaizaBrito:2007kz}.}. For the case in which $m\neq 0$ and holding the same conditions on the rest of the parameters we obtain a static configuration with a metric given by
\begin{equation}
ds^{2}=e^{2m}\left[d\tilde{t}^2-\frac{dr^2}{1+c_1/r^2}-r^{2}(d\theta ^{2}+\sin ^{2}{%
\theta }d\phi ^{2})\right]\,,
\end{equation}
wich scalar curvature reads
\begin{equation}
R=-2e^{-2m}\frac{c_1}{r^4}.
\end{equation}
Notice that this is just a conformal map of the above mentioned wormhole solution, with $\tilde{t}=t/\xi$.
\item
For $m=\gamma=0$, $\xi=1$ but $c_1$ being positive, the corresponding flux is complex. This means that from a string theory point of view, both fluxes RR and NS-NS are turned on. The metric Eq. (\ref{met}) reduces in this case to the one studied in \cite{Baekler:1987jb} which is given by
\begin{equation}
ds^2= dt^2-\frac{1}{1+c_1/r^2}dr^2-r^2(d\theta^2+\sin^2\theta d^2\phi),
\end{equation}
which corresponds to an exact solution of Einstein equations with a localized scalar field (Hodge dual to a 3-form flux) and a torsion kink. As it was shown in \cite{Baekler:1987jb}, these fields reside in an open einsteinian microcosmos. For $m\neq 0$ we obtain also an equivalent metric through the conformal factor $e^{2m}$. For vanishing and non-vanishing $m$, there is a naked singularity at $r=0$.

%
%
\end{enumerate}

\section{Conclusions and comments}

By using the group analysis method of differential equations, we have obtained an  exact evolving solution of the Einstein-Cartan equations of motion,  corresponding to a space-time threaded with a three-form Kalb-Ramond field strength.
The solution, related to the first invariant form constructed by this method, depends on four parameters namely $(m, \xi, \gamma, c_2)$, where $c_2$ measures the number of NS-NS flux present in the configuration.\\

This general dynamical solution shows some interesting features. First of all we observe that the corresponding scalar curvature vanishes for large time and large distances. This limit does not depend on the number  of flux we consider, which in the present case is only related to the the flux component ${\cal G}_{023}$ which can be real or complex. The extra components of the most generic flux ${\cal G}_3$ do not play a role in the solution. Moreover,  we see that for large distances and time running to large values, the scalar curvature reaches a zero constant value. The parameter $\xi$ does not play a role in the curvature values, which can be positive or negative according to the kind of fluxes, real or complex, giving rise to torsion. We also notice that as time runs to infinity, the curvature goes to zero even around the singularity at $r=0$. The corresponding space-time described by this solutions is highly curved around the singularity at $r=0$. However, as time runs, the curvature decreases and the space-time evolves into a soft universe with a null curvature for $r\neq 0$. Therefore, this space-time evolves into a an asymptotically flat space-time.\\

Special cases are commented: For instance, for the branch of solutions with $m=0$, there is one interesting case. The solution corresponding to the values $m=\gamma=0$ , $\xi=1$ and $c_1<0$
reduces to the static worm-hole solution found by
Sen Gupta et.al. in (\cite{6}). For generic values of parameters but $m=0$, we constructed a  solution which might be related to a dynamical evolving of a wormhole.  This is studied  in a forthcoming work. For $c_1>0$, the solution reduces to the spherically symmetric exact solution reported in \cite{Baekler:1987jb} which corresponds to an exact solution of Einstein equations with a localized scalar field (Hodge dual to a 3-form flux) and a torsion kink. Notice that $c_1$ can be positive or negative, according to which kind of fluxes we are considering. From the effective four-dimensional point of view, they are related to torsion components written in terms of real or complex three-form fluxes accordingly. From a string theory perspective, such fluxes correspond to type IIB string compactifications in the presence of NS-NS fluxes (positive $c_1$) or NS-NS and RR three-form fluxes (negative $c_1$).\\

On the other hand, solutions characterized by the value $\gamma=0$ represent static flux configurations. For the case in which the rest of parameters do not vanish, the solutions are asymptotically flat.\\

In the process, we have found evidence that there exist  more new solutions. Roughly speaking, the group analysis method consists on constructing invariant forms from which one could in principle, find an exact solution of the original set of differential equations. The first invariant form, as said, leads us to find the above described solution. For the second invariant form, the situation turned out to be more complicated. However, we were able to reduce an original set of five differential equations, into a set of only two different differential equations.  The solution is however difficult to find and we have left it for a future work. \\

\begin{center}
\textbf{ACKNOWLEDGMENTS}
\end{center}
We would like to thank Friedrich Hehl for useful suggestions and comments, and for pointing us out relevant references for our research. This work was partly supported by CONACyT Mexico, under contracts No. 49924-J, 105079 and 60209; also by SNI-CONACyT.  R.G-S. acknowledges partial support from COFAA, EDI, SIP 20100610 and
SIP 20110664 IPN grants. O. L.-B. is partially supported by  PROMEP.  F.F,  C.M. and A.q
Y.  acknowledges to ProSNI-UdG.

\end{document}